\documentclass[aps,pra,twocolumn,floatfix,superscriptaddress]{revtex4}

\usepackage{amssymb,amsmath,amstext}                
\usepackage{graphicx}                                               
\usepackage{epstopdf}                                               
\usepackage{color}                                                     
\usepackage{bm}                                                        
\usepackage{appendix}                                              
\usepackage[utf8]{inputenc}

\usepackage{ulem}
\usepackage{latexsym}
\usepackage[colorlinks=true,citecolor=blue,linkcolor=magenta]{hyperref}

\usepackage{mathtools}
\usepackage{float} 
\usepackage{amsthm} 

\DeclarePairedDelimiterX\braket[2]{\langle}{\rangle}{#1 \delimsize\vert #2} 

\def\be{\begin{equation}}
\def\ee{\end{equation}}
\def\bea{\begin{eqnarray}}
\def\eea{\end{eqnarray}}
\def\ra{\rangle}
\def\la{\langle}
\def\bi{\begin{itemize}}
\def\ei{\end{itemize}}
\def\ben{\begin{enumerate}}
\def\een{\end{enumerate}}

\definecolor{dgreen} {RGB}{78,138,21}

\newcommand{\as}[1]{{\color{blue}{{#1}}}}

\definecolor{dviolet} {RGB}{148,0,211}

\begin{document}

\title{Measurement of one-dimensional matter-wave quantum breather 
}

\author{Piotr Staro\'n} 
\affiliation{
Instytut Fizyki Teoretycznej, 
Uniwersytet Jagiello\'nski, ulica Profesora Stanis\l{}awa \L{}ojasiewicza 11, PL-30-348 Krak\'ow, Poland}
\author{Andrzej Syrwid} 
\affiliation{
Instytut Fizyki Teoretycznej, 
Uniwersytet Jagiello\'nski, ulica Profesora Stanis\l{}awa \L{}ojasiewicza 11, PL-30-348 Krak\'ow, Poland}
\author{Krzysztof Sacha} 
\affiliation{
Instytut Fizyki Teoretycznej, 
Uniwersytet Jagiello\'nski, ulica Profesora Stanis\l{}awa \L{}ojasiewicza 11, PL-30-348 Krak\'ow, Poland}
\affiliation{Mark Kac Complex Systems Research Center, Uniwersytet Jagiello\'nski, ulica Profesora Stanis\l{}awa \L{}ojasiewicza 11, PL-30-348 Krak\'ow, Poland
}

\begin{abstract}
Employing the Bethe ansatz approach and numerical simulations of measurements of particles' positions we investigate a post-quench many-body dynamics of attractively interacting bosons on a ring,  which in the mean-field approach corresponds to the so-called breather solution. Despite the fact that the initial many-body ground state is translationally invariant, the measurements reveal breather dynamics if quantum fluctuations of the center of mass of the system are extracted. Moreover, the analysis of the many-body evolution shows signatures of dissociation of the solitons that form the breather. 
\end{abstract}
\date{\today}

\maketitle

\section{Introduction}

Ultra-cold weakly interacting Bose gases can be successfully analyzed in the so-called mean-field approximation where all bosons are assumed to occupy the same single-particle state being a solution of the Gross-Pitaevskii equation (GPE) \cite{Pethick2002}.
The GPE is a non-linear equation that in the one-dimensional (1D) space can possess solitonic solutions \cite{Kivshar2011}. Interactions between ultra-cold atoms are usually described by the zero-range contact potential and if they are attractive the so-called bright soliton solution of the GPE representing the lowest mean-field energy state in the 1D space can appear \cite{Zakharov1971}.
For repulsive interactions the ground state corresponds to the uniform atomic density but there exist dark soliton solutions of the GPE which describe collectively excited Bose gas \cite{Zakharov1973}. 

When particles in a many-body system live in an infinite space or are confined in a ring geometry  the total momentum is a conserved quantum number and the many-body Hamiltonian of the system exhibits space translation symmetry. In such a case the probability density corresponding to the system eigenstates have to be also invariant under space translations of all particles by the same vector in space and thus the single-particle density calculated for an eigenstate has to be spatially uniform. On the other hand, it is clear that mean-field solitonic solutions break the space translation symmetry. 
Nevertheless, the existence of the solitonic solutions does not mean that the quantum many-body system has forgotten about the translation symmetry it should obey. There are quantum many-body processes, neglected in the mean-field description, which are trying to restore the symmetry \cite{McGuire1964,Calogero1975,Carter1987,Drummond1987,
Lai1989,Lai1989a,Rosenbluh1991,Drummond1993,
Corney1997,Corney2001,Carr2004,Mazets2006,
Calabrese2007,Muth2010,Bienias2011}. In the case of the bright soliton which describes the mean-field ground state of attractively interacting bosons, the soliton position coincides with the center of mass of the system. The center of mass can be initially prepared in a localized wavepacket but in the full many-body time evolution, the wavepacket starts spreading and after sufficiently long time we will not know where the soliton is located \cite{Castin_LesHouches,Castin2000,Syrwid2017}. Interestingly, the center of mass of a bright soliton can tunel through a potential barrier \cite{Weiss2009} or in the presence of a weak disorder potential it can Anderson localize \cite{Sacha2009,Delande2013}. In the dark soliton case quantum many-body fluctuations are a bit more difficult to describe because the position of the dark soliton is not the position of the center of mass of the system. Moreover, a dark soliton does not represent the ground state but a collectively excited state of the Bose system. Nevertheless, quantum many-body effects could be described \cite{Kulish1976,Ishikawa1980,Dziarmaga2002,Komineas2002,
Jackson2002,Dziarmaga2003,
Dziarmaga2004,Kanamoto2008,Dagnino2009,Kanamoto2010,
Martin2010,Karpiuk2012,Sato2012,
Delande2014a,Karpiuk2015,Kronke2015} and even the emergence of dark solitons in the course of measurements of positions of atoms prepared in translationally invariant many-body eigenstates was demonstrated \cite{Syrwid2015,Syrwid2016,Syrwid2017a,Syrwid2020,Oldziejewski2018,
Golletz2019}. 

Apart from the fundamental bright or dark solitons there are also higher order solitonic solutions of the GPE that describe, for instance, two solitons propagating in the 1D space that approach each other, collide and afterwards restore their initial shapes and propagate further \cite{Zakharov1971,Zakharov1973,Satsuma1974,Gordon1983}. 
While the parameters describing the positions and velocities of the individual solitons in the mean-field solutions can be the well defined classical parameters, in the many-body description of the Bose gas they are associated with quantum operators and consequently reveal quantum fluctuations. For example two dark solitons can stay at the same distance according to the mean-field predictions, while in the many-body description one observes quantum fluctuations of the relative distance between them \cite{Syrwid2015,Syrwid2016,Syrwid2017a,Syrwid2020}. It means that not only the location of an entire solitonic structure but also relative distances between solitons can be subject to quantum fluctuations due to many-body effects.

Another example of quantum fluctuations of a relative distance between solitons is  dissociation of the 1D breather which is a superposition of two bright solitons located at the same position \cite{Streltsov2008,Cosme2016,Weiss2016,Opanchuk2017,Yurovsky2017,
Marchukov2019a}. In the mean-field description the breather corresponds to periodic oscillations of the probability density due to periodic evolution of the relative phase between the two nonmoving solitons localized at the same position. In the full many-body approach, the relative position of the localized solitons reveals quantum fluctuations and neither the relative distance nor the relative velocity are well defined classical quantities and quantum dissociation of the breather is predicted \cite{Yurovsky2017,Marchukov2019a}. 

In the present paper  we are interested in quantum many-body effects that go beyond the mean-field approximation. On one hand, the total number of bosons $N$ has to be large enough if one wants to apply the mean-field approach. On the other hand, $N$ must be sufficiently small if one would like to observe quantum many-body effects in the laboratory otherwise time scale needed for the emergence of the effects is extremely long. In this article our analysis concentrates on the emergence of the breather from a translationally invariant many-body state of the Bose system and signatures of its dissociation. It turns out that even for a small number of bosons, the breather dynamics can be observed in the measurements of positions of particles provided fluctuations of the center of mass of the system, which on a ring is determined by the particles' barycenter, are extracted. 

\section{Model}

Ultra-cold bosonic atoms in the 1D space can be described by the Lieb-Liniger Hamiltonian \cite{Lieb1963, Lieb1963a},
\be
H=\sum_{i=1}^N\frac{p_i^2}{2}+\frac{g_0}{2}\sum_{i\ne j=1}^N\delta(x_i-x_j),
\label{MBH}
\ee
where units are chosen so that $\hbar=m=1$, $N$ is the total number of particles, $p_j=-i\partial_{x_j}$ and $g_0$ denotes the strength of the contact interactions which is determined by  the atomic $s$-wave scattering length. In the present paper we focus on the attractive interactions between atoms, i.e. $g_0<0$.

Let us start with the mean-field description of the system. If atoms form a Bose-Einstein condensate (BEC), the many-body state can be approximated by a product state where all bosons occupy the same single-particle wavefunction, i.e. $\Psi(x_1,x_2\dots,x_N,t)=\phi(x_1,t)\phi(x_2,t)\dots\phi(x_N,t)$, where $\phi(x,t)$ fulfills the Gross-Pitaevskii equation \cite{Pethick2002}
\be
i\partial_t\phi(x,t)=-\frac12 \partial_x^2\phi(x,t)-g|\phi(x,t)|^2\phi(x,t),
\label{GPE}
\ee
with $g=-g_0(N-1)$ and $\la\phi(t)|\phi(t)\ra=1$. Within the mean-field approximation, the ground state of the Bose system is described by the fundamental bright soliton solution 
\be
\phi(x,t)=\frac{\sqrt{g}\;e^{ig^2 t/8}}{2\cosh[g(x-x_{\rm cm})/2]}.
\label{fundamental}
\ee
It is a bound state of atoms which form a localized wavepacket located at $x_{\rm cm}$. The latter is the center of mass position of atoms which in the mean-field description is represented by a number --- a classical position variable. In the full many-body description, $x_{\rm cm}$ is a Hermitian operator and in the ground state the probability density to measure the center of mass position is uniform along the ring. However, the mean-field soliton density profile, Eq.~(\ref{fundamental}), emerges from the full many-body description if we calculate a particle density with respect to the center of mass position \cite{Castin_LesHouches,Castin2000}.

\begin{figure}[]  
\includegraphics[width=1\columnwidth]{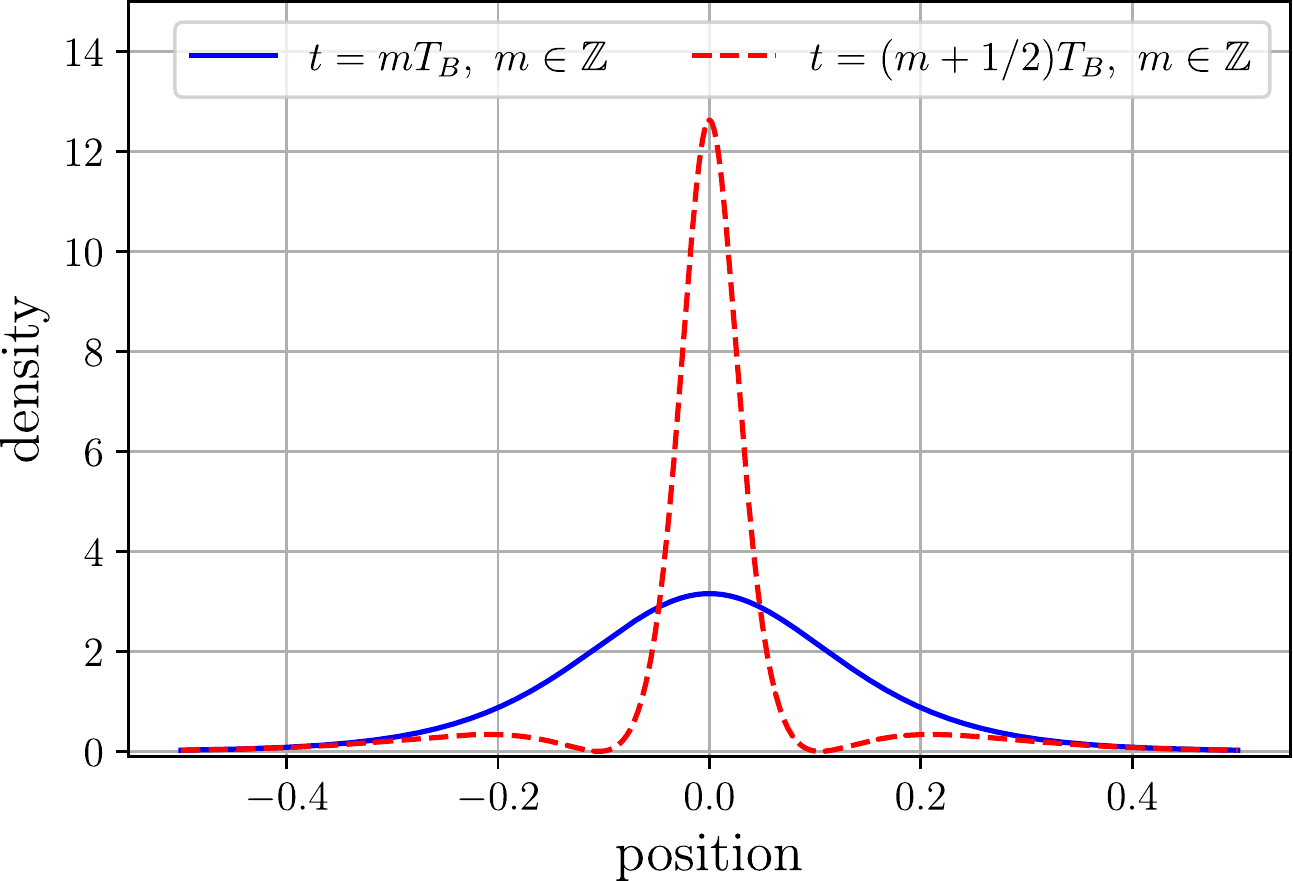}       
\caption{
Periodic evolution of the probability density corresponding to the mean-field breather solution, Eq.~(\ref{MF_breather}),  for $g=11.85$ and $x_{\mathrm{cm}}=0$. Note that  at $t=mT_B, \, m\in\mathbb{Z}$ (solid blue line) the mean-field breather solution coincides with the fundamental bright soliton, Eq.~(\ref{fundamental}). After half of the period, i.e. at $t=\left(m+\frac{1}{2}\right)T_B$, we can observe a very high and narrow central peak accompanied by two small side maxima located symmetrically on the left and on the right (dashed red line).
}
\label{mean_breather_den}   
\end{figure}

Apart from the fundamental bright soliton, there are also higher order solitonic solutions of the GPE. In the present paper we consider the so-called breather solution. Suppose that we have prepared the system in the mean-field product state with $\phi$ like in Eq.~(\ref{fundamental}) but at $t=0$ the interaction strength in the GPE is suddenly increased by a factor four, i.e. $g\rightarrow 4g$. Then, $\phi_{\rm B}(x,0)=\phi(x,0)$ but for $t>0$ the mean-field time evolution of the system is the following 
\be
\phi_{\rm B}(x,t)=\frac{\sqrt{g}\cosh(3\tilde x)+3\sqrt{g}\cosh(\tilde x)e^{i\omega_{\rm B} t}}{3\cos\omega_{\rm B} t+4\cosh\left(2\tilde x\right)+\cosh\left(4\tilde x\right)}
e^{i \omega_{\rm B} t/8},
\label{MF_breather}
\ee
where $\tilde x=g(x-x_{\rm cm})/2$ and $\omega_{\rm B}= g^2$. Equation~(\ref{MF_breather}) describes two bright solitons localized at $x_{\rm cm}$ whose relative phase changes with the period $T_{\rm B}=2\pi/\omega_{\rm B}$, see the phase factor $e^{i\omega_{\rm B}t}$ in the numerator of Eq.~(\ref{MF_breather}). Both solitons do not move and the resulting probability density oscillates as depicted in Fig.~\ref{mean_breather_den}.  It is a special case of a more general solution describing two solitons with a mass ratio $3:1$ where these solitons can be localized at different positions in space or propagate with different constant velocities \cite{Gordon1983}.  At time moments equal to integer multiple of $T_{\rm B}$ the probability density $|\phi_{\rm B}(x,t)|^2$ matches the probability density of the fundamental bright soliton, Eq.~(\ref{fundamental}). 

Now let us switch to the full many-body description within the Bethe ansatz approach \cite{Lieb1963, Lieb1963a,korepin_bogoliubov_izergin_1993,Gaudin2014_book,Syrwid2020}. In the case of the contact interactions, atoms behave like free particles except moments when they collide. Thus, if positions of all particles are different, many-body eigenstates of the system must reduce to plane waves $e^{ik_1x_1}e^{ik_2x_2}\dots e^{ik_Nx_N}$. When two particles meet each other (i.e. $x_i=x_j$) eigenstates have to fulfill the boundary conditions determined by the strength $g_0$ of the contact interactions  \cite{Gaudin2014_book}. Moreover, states of a Bose system must be symmetric with respect to an exchange of any two bosons. It results in many-body eigenstates which are superpositions of $N!$ terms of plane waves. Assuming that atoms are on a ring with the circumference $L$ and fulfill the periodic boundary conditions, the parameters $k_j$ of the plane waves, which are called quasi-momenta, are solutions of the Bethe equations \cite{Lieb1963,Lieb1963a},
\be
e^{ik_j L} = - \prod_{s=1}^{N} \frac{k_j-k_s+ig_0}{k_j-k_s-ig_0}, \hspace{0.75cm} j=1,2,\ldots,N.
\label{BEQ}
\ee
The set of quasimomenta $k_{j=1,\ldots,N}$ satisfying Eqs.~(\ref{BEQ}) determines an eigenstate which is characterized by the eigenenergy $E=\frac{1}{2}\sum_{j=1}^Nk_j^2$ and the total momentum $P=\sum_{j=1}^Nk_j$. 

The translation symmetry of the many-body Hamiltonian in Eq.~(\ref{MBH}), resulting in the total momentum conservation, implies that the density of atoms (i.e. the single-particle probability density multiplied by $N$) corresponding to an eigenstate must be uniform in space. One may ask where is the fundamental bright soliton, Eq.~(\ref{fundamental}), that in the mean-field approach describes the ground state of the attractively interacting bosons. It turns out that even if the system is prepared in the translationally symmetric many-body ground state (i.e. an eigenstate of the translation operator that shifts positions of all particles by the same distance), the soliton will emerge in the measurements of particles' positions but we do not know where on the ring it will turn up \cite{Castin_LesHouches,Castin2000,Syrwid2017}. Indeed, in different realizations of the same experiment, the measurements of the atomic density will result in the profile very well approximated by the probability density corresponding to the state in Eq.~(\ref{fundamental}). However, in each realization $x_{\rm cm}$ will appear as a random number chosen with the uniform distribution on the ring. 

In the following we analyze the emergence of the breather dynamics when the Bose system is initially prepared in the translationally invariant many-body ground state and the interaction strength is quenched by a factor of four, i.e. $g_0\rightarrow 4g_0$.

\section{Results}

Exact many-body simulations of the Bose system described by the Lieb-Liniger Hamiltonian in Eq.~(\ref{MBH}), within the Bethe ansatz approach, are possible for a small number of particles only. This is due to dramatic proliferation of the Bethe eigenstates complexity with increasing $N$ \cite{Lieb1963,Lieb1963a,korepin_bogoliubov_izergin_1993,Gaudin2014_book,Syrwid2020}. Moreover, in order to obtain eigenstates of the system, the non-linear Bethe equations have to be solved numerically (only in the limiting cases of very weak or very strong interactions approximate analytical solutions are attainable). For attractive interactions, even numerically it is not easy to get solutions of the Bethe equations because the so-called quasi-momenta are complex-valued and extremely high numerical precision is required in order to obtain eigenstates which fulfill periodic boundary conditions. In addition, in comparison to the ground state the numerical determination of quasi-momenta corresponding to excited eigenstates, that may undergo bifurcation with the change of the attraction strength, is even more challenging. Therefore, when we analyze time evolution of the system after the quench of the interactions we restrict to $N=4$ (see also \cite{Zill2018} for time evolution of a similar system). In the case of the fundamental bright soliton it was shown that $N\gtrsim 3$ is sufficient to observe the mean-field behavior in the quantum many-body description \cite{Mazets2006}. In the previous study of the breather \cite{Weiss2016}, for such a small particle number, signatures of the mean-field evolution could not be identified in the many-body simulations where the Lieb-Liniger model was approximated by the Bose-Hubbard Hamiltonian. 

Exact many-body analysis predicts dissociation of the two solitons that form the breather but the picture how the breather dynamics and its dissociation look like in the measurements of the atomic density has not been demonstrated within the Bethe ansatz formalism \cite{Yurovsky2017}. In the large number of particles limit, one may apply approximate many-body methods and it was shown that the particle density of the Bose system prepared initially in a localized state reveals the breather oscillations which decay in time due to quantum many-body effects \cite{Opanchuk2017,Yurovsky2017}. The decay of the oscillations are also visible in the second order correlation function indicating that not only the center of mass of the system but also the relative distribution of particles is spreading in time \cite{Opanchuk2017}.

Here we focus on the exact description of the small Bose system prepared initially in a translationally invariant state and show that the breather dynamics can be observed if quantum fluctuations of the center of mass position are extracted. In addition, a careful analysis of the many-body breather evolution allows us to identify signatures of the dissociation process.

\begin{figure}
\center         
\includegraphics[width=1\columnwidth]{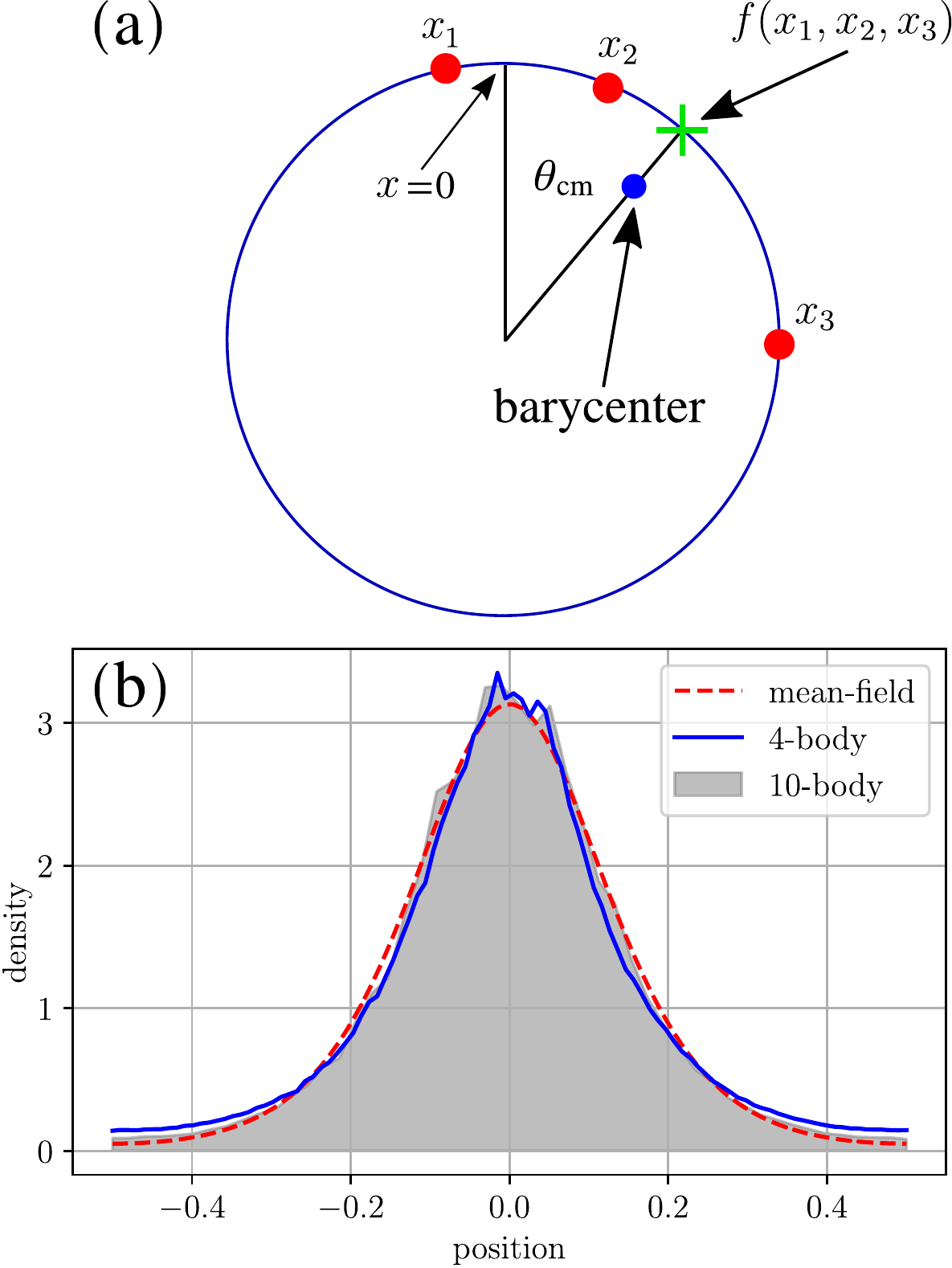}
\caption{Panel (a) illustrates how the center of mass of particles on a ring is determined in the three-particle case. First the barycenter in the 2D space is calculated  and next the line starting from the center of the ring and passing through the barycenter is drawn and its crossing with the ring determines the center of mass position of particles. Note that the barycenter is associated with the angle $\theta_\mathrm{cm}$, which measures the angular distance from the reference position $x=0$. Panel~(b): the normalized histogram (solid blue line) presents the single-particle probability density $\rho_{\rm rel}(x)$, Eq.~(\ref{rho1mod}), where $\Psi=\Psi_0$ is the many-body ground state of $N=4$ particles which attract each other with the strength $g_0(N-1)=-11.85$. The histogram consists of $10^6$ realizations of the measurements of the positions of $N=4$ particles, see Appendix~\ref{BB}. Similar result but obtained for $N=10$ is illustrated by a gray shading. For comparison, the dashed red line shows the probability density of the corresponding mean-field ground state which is the fundamental bright soliton, Eq.~(\ref{fundamental}).}
\label{bary_bright}   
\end{figure} 

Let us consider $N=4$ atoms on a ring of the circumference $L=1$. The particles are initially prepared in the ground state $\Psi_0(x_1,\dots,x_N)$ of the Lieb-Liniger Hamiltonian, Eq.~(\ref{MBH})\as{,} where the initial interaction strength $g_0 = -3.95$. In the GPE, Eq.~(\ref{GPE}), the parameter $g=-g_0(N-1)= 11.85$ what implies that the attractive interactions are sufficiently strong (i.e. $g>\pi^2$) for a bright soliton to form in a ring geometry. It is worth stressing that with increasing $g$, the mean-field bright soliton on a ring quickly approaches the fundamental bright soliton solution in the infinite space, Eq.~(\ref{fundamental}). In the many-body description, the ground state is translationally invariant and the corresponding single-particle probability density, 
\be
\rho(x)=\int_{-1/2}^{1/2}|\Psi(x_1,\dots,x_{N-1},x)|^2dx_1\dots dx_{N-1},
\label{rho1}
\ee
is spatially uniform, i.e. $\rho(x)=1$ for $\Psi=\Psi_0$. 

For $N=4$ the mean-field bright soliton profile can only emerge if we measure positions of atoms in many realizations of the same experiment and prepare the histogram of the particles' positions. Before the histogram is prepared, the detected positions have to be shifted so that the center of mass of the system on a ring is always located at the same point in each realization of the same experiment \cite{Castin_LesHouches,Castin2000,Syrwid2017,Syrwid2020}. 
In the case of particles confined in  a ring geometry, it is not straightforward to determine the center of mass position. 
In order to do that we first calculate the barycenter of particles in a 2D plane which contains the ring and then determine the angle $\theta_{\mathrm{cm}}$ corresponding to the center of mass position on a ring $\frac{\theta_\mathrm{cm}}{2\pi}L\equiv f$, see Fig.~\ref{bary_bright}(a). The bright soliton profile can be obtained by calculating the single-particle probability density like in Eq.~(\ref{rho1}) provided the position of the $N$-th particle is shifted with respect to the center of mass position $f(x_1,\ldots,x_{N-1})$ of the remaining $N-1$ particles, 
\bea
\rho_{\rm rel}(x)&=&\int_{-1/2}^{1/2}\big|\Psi[x_1,\dots,x_{N-1},x-f(x_1,\dots,x_{N-1})]\big|^2
\cr && \times dx_1\dots dx_{N-1}.
\label{rho1mod}
\eea
Note that due to the indistinguishability of bosons it does not matter which particle is chosen as the $N$-th particle. The integral in Eq.~(\ref{rho1mod}) is calculated by means of the Monte Carlo integration (see Appendix~\ref{BB}) and the result for $\Psi=\Psi_0$ are presented in Fig.~\ref{bary_bright}(b) together with the mean-field solution for the fundamental bright soliton on a ring. 

In the mean-field case, if the fundamental bright soliton is chosen as the initial state and the interaction strength is quenched by a factor of four, $g\rightarrow 4g$, the breather dynamics is observed, cf. Eq.~(\ref{MF_breather}). In the many-body description, we choose the 4-particle ground state of the system as the initial state,
\bea
\Psi(x_1,x_2,x_3,x_4;t=0)=\Psi_0(x_1,x_2,x_3,x_4),
\label{psi_init}
\eea
and we would like to examine if the signatures of the breather can be observed when the interactions are quenched similarly to the mean-field case. To perform the time evolution, the wavefunction in Eq.~(\ref{psi_init}) is expanded in the basis of the 4-particle eigenstates of the Lieb-Liniger Hamiltonian after the quench. The 4-particle eigenstates are obtained by means of the Bethe ansatz approach and the projections of $\Psi(x_1,x_2,x_3,x_4;t=0)$ on the eigenstates after the quench are determined by analytical 4-dimensional integrations, see Appendix~\ref{BB}. 

\begin{figure}[h!]         
\includegraphics[width=1\columnwidth]{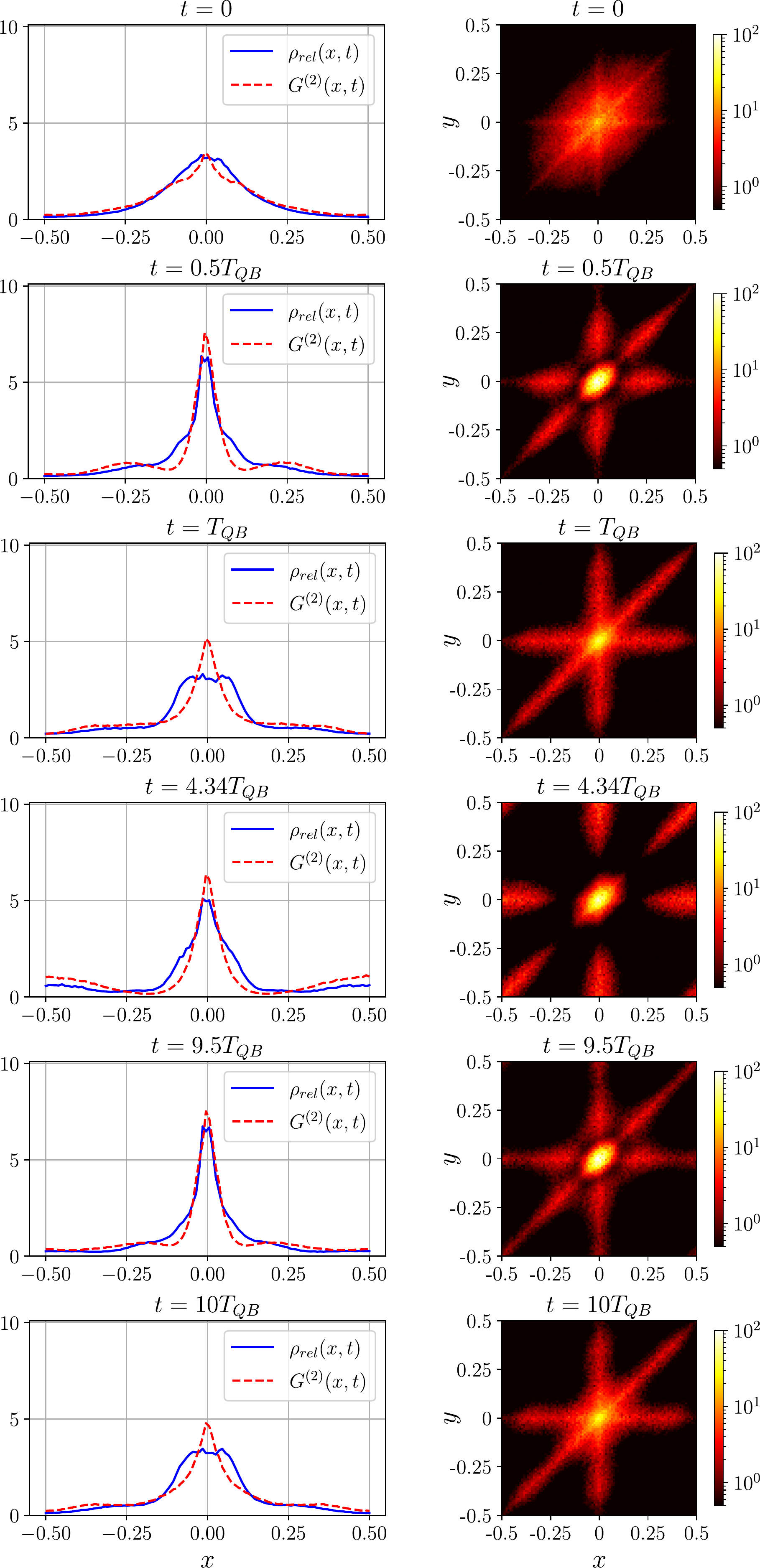}    
\caption{
Time evolution of the 4-particle system. The system is initially prepared in the ground state for $g_0=-3.95$ and then the interactions are quenched by a factor of four, i.e. $g_0\rightarrow 4g_0$.  Left column shows $\rho_{\rm rel}(x;t)$, Eq.~(\ref{rho1mod}), (solid blue line) and $G^{(2)}(x;t)$, Eq.~(\ref{g2}), (dashed red line) for different moments of time as indicated above the panels. Right column: color coded plots of $P(x,y;t)$, Eq.~(\ref{Pxyt}), at the same time moments as in the corresponding left panels. By monitoring the time evolution of the considered system we estimate the period of the quantum breather oscillations $T_{QB}\approx 0.0168$. At $t_d\approx 0.073$ the oscillations nearly die out, cf. Fig.~\ref{PeakOsc} where the time moments presented in the current figure are indicated by vertical black lines.
}\vspace{-0.3cm}
\label{densities}   
\end{figure}

The modulus of the wavefunction $\Psi(x_1,x_2,x_3,x_4;t)$ is translationally invariant at any time. 
Thus, when we plot the single-particle probability density $\rho(x;t)$, Eq.~(\ref{rho1}), we always obtain a uniform distribution which reflects the fact that the center of mass of the system is perfectly delocalized in a translationally symmetric state. However, if the fluctuations of the center of mass position are extracted, signatures of the breather dynamics emerge. Indeed, left column of Fig.~\ref{densities} shows that up to $t=t_d\approx0.073$,the time evolution of the single-particle distribution $\rho_{\rm rel}(x;t)$, Eq.~(\ref{rho1mod}), reveals strong oscillations of the central peak which resemble the mean-field behavior of the breather. \ This observation is further supported by the behavior of the second order correlation function,
\be
G^{(2)}(x-y;t)=\int_{-1/2}^{1/2} |\Psi(x,y,x_3,x_4;t)|^2dx_3 dx_4,
\label{g2}
\ee
which, apart from the oscillations of the central peak, reveals also a periodic appearance of two side maxima, similar to those which characterize the mean-field breather dynamics, cf. Figs.~\ref{mean_breather_den} and \ref{densities}.
Note that for a translationally invariant state, $G^{(2)}$  depends only on the relative distance $x-y$ between two particles and is completely unaffected by fluctuations of the center of mass position. Figure~\ref{densities} shows that for $t\lesssim t_d$, according to  $G^{(2)}(x-y;t)$, two particles prefer to locate close to each other but there appear also maxima at $|x-y|\approx0.25$. These side maxima are deformed and thus not clearly visible in $\rho_{\rm rel}(x;t)$ due to a small number of particles $N-1=3$ used to determine the center of mass position $f(x_1,x_2,x_3)$. If we assume that three particles are measured at the same position, e.g. at $x=0$, then the probability density of the wavefunction for the fourth particle $\psi(x_4)=\Psi(0,0,0,x_4,t)$ shows clearly the side maxima and its phase reproduces the behavior of the phase of the mean-field breather solution. Note that the period of the observed oscillations $T_{QB}\approx 0.0168$ is different from the mean-field prediction for the breather which is equal to $T_B\approx0.0447$ but we should not expect quantitative agreement for the $N=4$ particle system.

\begin{figure}[h!]         
\includegraphics[width=1.\columnwidth]{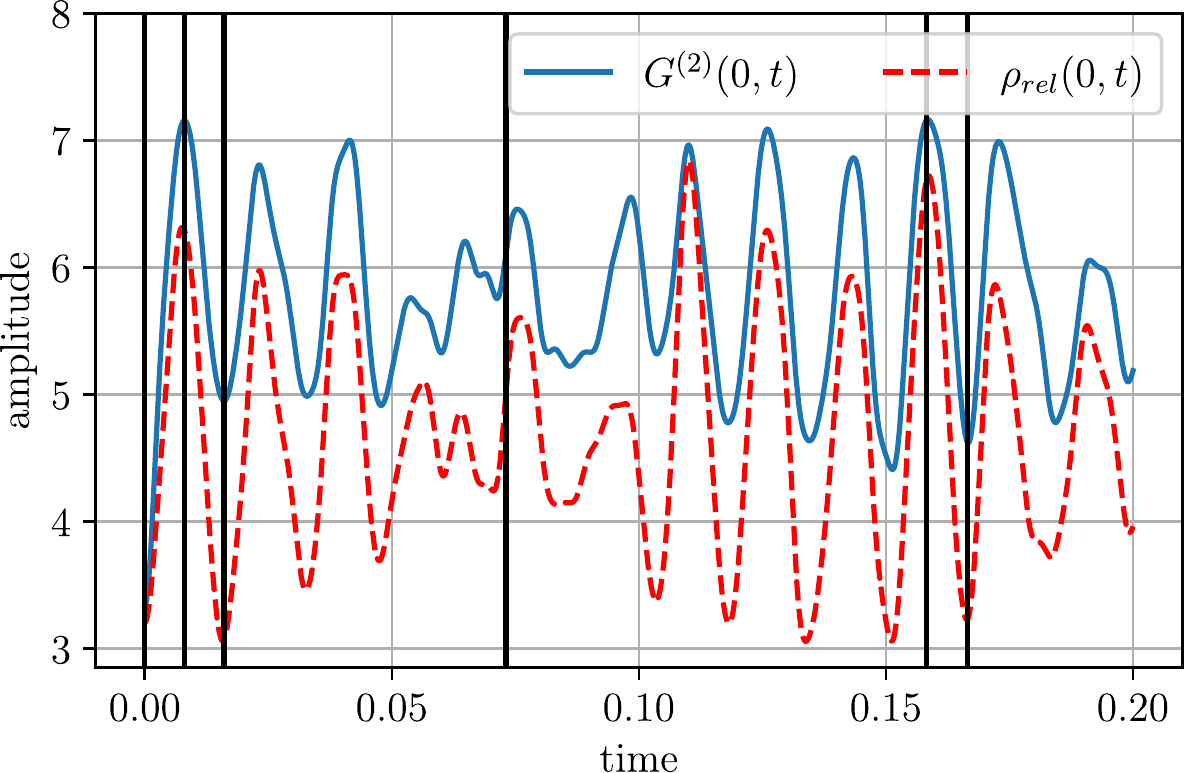}      
\caption{Time evolution of $\rho_{\rm rel}(0;t)$ (dashed red line) and $G^{(2)}(0;t)$ (solid blue line) which corresponds to the results presented in Fig.~\ref{densities} and illustrates decay and revival of the amplitudes of the central peak oscillations in the plots of $\rho_{\rm rel}(x;t)$ and $G^{(2)}(x;t)$. While initially the oscillations are significant and indicate the breather-like dynamics known from the mean-field description, for longer times (of the order of $t_d\approx 0.073$) they nearly die out. The decay of the central peak oscillations is accompanied by the increasing distance between the central peak and side maxima, cf. Fig.~\ref{densities}.}
\label{PeakOsc}   
\end{figure}

In Fig.~\ref{PeakOsc} we present $\rho_{\rm rel}(0;t)$ and $G^{(2)}(0;t)$, i.e. the temporal behavior of the amplitudes of the central peak oscillations in the single-particle density $\rho_{\rm rel}$ and in the correlation function $G^{(2)}$. It turns out that the oscillations of the central peak nearly die out at $t\approx t_d$ and consequently signatures of the breather dynamics are not observed. In addition, analyzing the plots of $G^{(2)}(x-y;t)$, one can observe that the damping of the central peak oscillations is accompanied by an increasing distance between the central peak and the side maxima, compare Figs.~\ref{densities} and \ref{PeakOsc}. This can be attributed to dissociation of the breather where the big soliton consisting of $3N/4$ atoms and the small soliton that contains $N/3$ atoms are supposed to move apart \cite{Yurovsky2017}. One should keep in mind that the quantum state $\Psi$ does not favour the situation when the big soliton is moving to the right and the small soliton towards the left over the situation when the directions of their motion are reversed. In other words signatures of the both scenarios are present in many realizations of particles' positions measurements. In the result the plots of the average density and the correlation function exhibit the reflection symmetry with respect to the position of the central peak. We interpret the increasing distance between the central peak and the side maxima, accompanied by the damping of the central peak oscillations,  as signatures of the breather dissociation. These features are better visible in the plots of the conditional probability for the detection of two particles provided one particle is initially measured at a fixed position, e.g. at $t=0$ the third particle is measured at $x_3=0$, 
\be
P(x,y;t)=\int_{-1/2}^{1/2}|\Psi(x,y,0,x_4;t)|^2d x_4.
\label{Pxyt}
\ee
Figure~\ref{densities}  shows that two particles most probably can be detected at the same position as the third one (i.e. at $x\approx y\approx x_3 = 0 $) or around six side maxima visible in the color coded plots of $P(x,y;t)$. Note that the distance between the side maxima and the central peak in $P(x,y;t)$ increases in time reaching a maximal value at $t_d\approx0.073$ when the amplitudes of the oscillations of $\rho_{\rm rel}(0;t)$ and $G^{(2)}(0;t)$ are minimal. We can also calculate the time needed for two dissociating solitons to get on the opposite side of the ring. Assuming that their relative velocity is approximated by the velocity scale $v_{0}$ of dissociating solitons estimated in Ref.~\cite{Yurovsky2017}, i.e. $v_{0} = 2 \as{|}g_0\as{|}= 7.9$, we obtain  $t\approx0.063$ which is comparable to $t_d$.  For longer time evolution, the $N=4$ particle system shows quantum revival where the breather dynamics that we have observed initially returns but not ideally.

The many-body state $\Psi(x_1,x_2,x_3,x_4;t)$ does not describe a BEC with all bosons occupying the single-particle wavefunction corresponding to the mean-field breather solution, Eq.~(\ref{MF_breather}). The translationally invariant state $\Psi$ can be rather interpreted as a superposition of two-soliton solutions with different positions and velocities of the solitons \cite{Yurovsky2017}. In the case of a fundamental bright soliton or a dark soliton, a single realization of the measurement of particles' positions reveals clearly a soliton but its location is random in the different realizations due to strong quantum fluctuations of the center of mass of the system \cite{Castin_LesHouches,Syrwid2015,Syrwid2016,Syrwid2017,Syrwid2020}. In the present case, the situation is more complicated because averaged results of the measurements consist of mixtures of different mean-field two-soliton contributions. In addition, the size of the investigated structures is not much smaller than the ring circumference and thus in the time evolution they feel periodic boundaries quite quickly which also distorts the results. Consequently, we cannot expect clear pictures of the breather dynamics or the dissociation of the solitons but only signatures of them. Moreover, the system we analyze in the present paper is small and in order to plot density of atoms we have to perform simulations of the detection of particles' positions many times. If a large system was attainable numerically, then single realizations of the measurement process would allow for the plots of the densities and probably different solitonic structures would be much better visible.

\section{Summary and conclusions}

Bose systems which form Bose-Einstein condensates can be quite accurately described by the Gross-Pitaevskii equation which in the 1D case can possess soliton solutions. The latter do not obey the space translation symmetry despite the fact that the original many-body Hamiltonian is translationally invariant. In such circumstances one may expect quantum many-body effects which go beyond the mean-field approximation and which can be responsible for destruction of solitons or quantum fluctuations of solitons' parameters. 

In the present paper we investigate a quantum many-body system that in the mean-field approximation is described by a higher order soliton solution called the breather. Apart from quantum fluctuations of the center of mass of the system, which have been already investigated in the case of single soliton solutions \cite{Castin_LesHouches,Syrwid2015,Syrwid2016,Syrwid2020}, there are additional degrees of freedom that can suffer from quantum many-body effects. That is, the relative position of two solitons, which form the breather, and their relative velocity reveal quantum fluctuations. Such fluctuations are expected to lead to the breather dissociation \cite{Yurovsky2017,Marchukov2019a} 

Here we show that the breather dynamics can emerge from the translationally invariant ground state of the system after the interactions are quenched if one performs measurements of particles' positions. Moreover, time evolution of the quantum many-body system reveals signatures of the breather dissociation where the two solitons are moving away from each other. 

The exact many-body approach we perform allows for the analysis of a small system only. There are approximate many-body methods that are valid for a large particle number $N$ which were already applied to the breather problem \cite{Yurovsky2017,Opanchuk2017,Marchukov2019a}. It would be very interesting if the gap between the small and large values of $N$ was filled. 
It is the regime where quantum many-body fluctuations are expected to have strong influence on the system behavior on the time scale attainable experimentally. Moreover, soliton structures, that could emerge in the measurement process, should be in quantitative agreement with the mean-field predictions.

\section*{Acknowledgements}

The authors would like to express their sincere gratitude to Vladimir A. Yurovsky for drawing their attention to the quantum breather problem and for fruitful discussions.
Support of the National Science Centre, Poland via Projects No.~2016/21/B/ST2/01086 (P.S.), No.~2018/28/T/ST2/00372 (A.S.), and  No.~2018/31/B/ST2/00349
(K.S)  is acknowledged. A.S. acknowledges the support of the Foundation for Polish Science (FNP). 

\appendix
\section{Eigenstates of the Lieb-Liniger model within the Bethe ansatz approach}
\label{AA}
Eigenstates of the $N$-particle Lieb-Liniger Hamiltonian, Eq.~(\ref{MBH}), can be cast into the following form \cite{Lieb1963, Lieb1963a,korepin_bogoliubov_izergin_1993,Gaudin2014_book}
\begin{equation}
	\Psi_{\{k\}}(\{x\}) 
	  = \mathcal{N} \!\!\sum_{\sigma\in\mathcal{S}_N}\!\! \mathcal{A}_\sigma(\{x\},\{k\}) \exp\!\left(i \sum_{j=1}^N k_{\sigma(j)} x_j \right), 
\label{eqn::Bethe}
\end{equation}
where  $\mathcal{S}_N$ is the group of all $N$-element permutations,
\begin{equation}
	 \mathcal{A}_\sigma(\{x\},\{k\})=\prod_{l>m}{\left(1 - \frac{i \ g_0 \, \mathrm{sgn}(x_l-x_m)}{k_{\sigma(l)}-k_{\sigma(m)}}\right)}, 
\label{eqn::Bethe2}
\end{equation}
and the normalization constant
\begin{equation}
\mathcal{N}= \frac{\prod_{l>m}{(k_l-k_m)}}{\sqrt{N!\,  \det[\mathcal{M}({k})] \, \prod_{l>m}{\big[(k_l-k_m)^2 + g_0^2\big]  }}},
\label{normApp}
\end{equation}
can be determined with the help of the Gaudin matrix
\begin{equation}
\mathcal{M}_{\alpha\beta} = \delta_{\alpha \beta} \left(1+\sum_{l=1}^N \frac{2 g_0}{(k_{\alpha} - k_l)^2+g_0^2}\right) - \frac{2 g_0}{(k_{\alpha} - k_{\beta})^2+g_0^2}.
\end{equation}
In the case of periodic boundary conditions the set of $N$ quasimomenta $\{k\}$ has to satisfy the Bethe equations, Eqs.~(\ref{BEQ}), and can be found iteratively starting from the weakly interacting limit $|g_0|L\ll 1$ where the Bethe equations can be approximated by 
\begin{equation}
k_j =\frac{2\pi}{L}d_j+\frac{g_0}{L} \sum_{\substack{s=1 \\ s\neq j}}^N\frac{1}{k_j-k_s}, \hspace{0.8cm} j=1,2,\ldots, N.
\label{ApproxBethe}
\end{equation}
The numbers $d_j=0,\pm 1, \pm 2, \ldots$ do not have to be distinct for different $j$ and they denote excitations if $|d_j|\neq 0$ \cite{Batchelor2004}. That is, in the weakly interacting limit the set of quasimomenta determining the ground state can be found by setting $d_{j=1,2,\ldots,N}=0$. Other sets $\{k\}$ of approximated values of quasimomenta corresponding to excited eigenstates are related to at least one  $d_j$ different from zero. The solutions of the approximate Eqs.~(\ref{ApproxBethe}) can be used as an initial guess to solve the original Bethe Eqs.~(\ref{BEQ}).  In the consecutive steps we slightly increase the coupling strength determined by $g_0$ and use the linear extrapolation basing on the previous solutions of Eqs.~(\ref{BEQ}). Starting from different sets $\{d_1,d_2,\ldots,d_N\}$, excluding permutations of $d_j$ numbers, we obtain different solutions of the Bethe Eqs.~(\ref{BEQ}). For more details see Ref.~\cite{Sykes}.

\section{Determination of overlaps between different eigenstates}
\label{BB}

Let us consider two different $N$-particle eigenstates $\Psi_{\{k\}}(\{x\})$ and $\widetilde{\Psi}_{\{q\}}(\{x\})$, Eq.~(\ref{eqn::Bethe}), but corresponding to the Lieb-Liniger systems with possibly different coupling strengths determined by $g_0$ and $\widetilde{g}_0$, respectively. It turns out that the overlap between these two eigenstates can be written as the following sum of $(N!)^2$ integrals
\begin{equation}
\begin{split}
	& \big<\widetilde{\Psi}\big|\Psi \big> = \mathcal{N} \mathcal{\widetilde{N}}^*  N! \sum_{\pi \in \mathcal{S}_N}\sum_{\sigma \in \mathcal{S}_N} \widetilde{A}_\sigma^*(\{q\}) \, A_\pi(\{k\})   \\
 & \times \int\limits_0^L \!dx_L \int\limits_{x_1}^L \!dx_2 \ldots \int\limits_{x_{N-1}}^L \!\! dx_N \, \exp\!\left(i \sum_{j=1}^N (k_{\pi(j)} - q^*_{\sigma(j)})x_j \right),
\end{split}
\label{eqn::overlap}
\end{equation}
where 
\begin{equation}
	 A_\pi(\{k\})=\prod_{l>m}{\left(1 - \frac{i  g_0  }{k_{\pi(l)}-k_{\pi(m)}}\right)},
\label{eqn::overlap2}
\end{equation}
\begin{equation}
	 \widetilde{A}_\sigma^*(\{q\})=\prod_{l>m}{\left(1 + \frac{i  \widetilde{g}_0  }{q^*_{\sigma(l)}-q^*_{\sigma(m)}}\right)},
\label{eqn::overlap3}
\end{equation}
and $\widetilde{\mathcal{N}}$ is given by Eq.~(\ref{normApp}) with $\{k\}$ and $g_0$ replaced by $\{q\}$ and $\widetilde{g}_0$, respectively.
Note that for small number of particles $N$, like $N=4$ which we analyze in the main text, the overlap given by the analytical expression~(\ref{eqn::overlap}) can be explicitly calculated.

\begin{proof}[Proof of Eq.~(\ref{eqn::overlap})]
	The overlap in question reads 
\begin{equation}
\big<\widetilde{\Psi}\big|\Psi \big> = 
 \int\limits_0^L \!dx_1 \!\int\limits_{0}^L \!dx_2 \ldots \!\int\limits_{0}^L \!dx_N\, \widetilde{\Psi}^*_{\{q\}}(\{x\}) \Psi_{\{k\}}(\{x\}).
 \label{defapp}
\end{equation}
First of all, we observe that the integration over the $N$-dimensional hypercube can be decomposed into a sum of $N!$ integrations over the sectors in which $0 \leq x_{\tau(1)} \leq x_{\tau(2)} \ldots\leq x_{\tau(n)} \leq L$, where $\tau$ is the permutation belonging to $ \mathcal{S}_N$, i.e.
\begin{equation}
\begin{split}
 \int\limits_0^L &  dx_1 \int\limits_{0}^L dx_2 \ldots \int\limits_{0}^L dx_n \\
  & = \sum_{\tau \in \mathcal{S}_N} \int\limits_0^L dx_{\tau(1)} \int\limits_{x_{\tau(1)}}^L \!dx_{\tau(2)} \ldots \!\int\limits_{x_{\tau(N-1)}}^L \!\!dx_{\tau(N)}.
 \end{split}
 \label{decompos}
\end{equation}
Note that by employing such a decomposition we get rid of the sign function present in the coefficients ${\cal A}$  [Eq.~(\ref{eqn::Bethe2})]. The latter become the coordinate independent numbers $A$ defined in Eq.~(\ref{eqn::overlap2}). Moreover, thanks to the Bose exchange symmetry, i.e. $\Psi_{\{k\}}(\tau\{x\})=\Psi_{\{k\}}(\{x\})$ and similarly $\widetilde{\Psi}_{\{q\}}(\tau\{x\})=\widetilde{\Psi}_{\{q\}}(\{x\})$, the analyzed expression can be reduced as follows
\begin{equation}
\begin{split}
 & \sum_{\tau \in \mathcal{S}_N} \int\limits_0^L \!dx_{\tau(1)} \ldots \!\! \int\limits_{x_{\tau(N-1)}}^L \!\!dx_{\tau(N)} \,\widetilde{\Psi}^*_{\{q\}}(\{x\}) \Psi_{\{k\}}(\{x\})  \\
 &=\sum_{\tau \in \mathcal{S}_N} \int\limits_0^L \!dx_{1}\ldots \!\! \int\limits_{x_{N-1}}^L \!\! dx_{N} \,\widetilde{\Psi}^*_{\{q\}}(\tau^{-1}\{x\}) \Psi_{\{k\}}(\tau^{-1}\{x\})  \\
 &=\sum_{\tau \in \mathcal{S}_N} \int\limits_0^L \!dx_{1}\ldots \!\! \int\limits_{x_{N-1}}^L \!\! dx_{N}\,\widetilde{\Psi}^*_{\{q\}}(\{x\}) \Psi_{\{k\}}(\{x\}) \\
  &= N!  \int\limits_0^L \!dx_{1}\ldots \!\! \int\limits_{x_{N-1}}^L \!\! dx_{N}\,\widetilde{\Psi}^*_{\{q\}}(\{x\}) \Psi_{\{k\}}(\{x\}).
 \end{split}
 \label{eqn::overlap_proof}
\end{equation}
Thus, Eq.~(\ref{eqn::overlap}) can be easily reproduced by employing Eqs.~(\ref{decompos})--(\ref{eqn::overlap_proof}) in Eq.~(\ref{defapp}). 
\end{proof}

In the present paper we analyze $N=4$ bosons prepared initially in the ground state, i.e. $\Psi(t=0)=\Psi_0$, for a certain value of the  interaction strength $g_0$. In order to obtain time evolution of the many-body state $\Psi(t)$ after the quench of the interactions, i.e. when $g_0\rightarrow \tilde g_0=4g_0$, we have to calculate overlaps $\alpha_{\{q\}}=\big<\widetilde{\Psi}_{\{q\}}\big|\Psi_0 \big>$ where $\widetilde{\Psi}_{\{q\}}$'s are eigenstates corresponding to $\tilde g_0$. Then,
\begin{equation}
	\Psi(\{x\},t) = \sum_{\{q\}} \alpha_{\{q\}}\, e^{-i E_{\{q\}} t} \,\,\widetilde{\Psi}_{\{q\}}(\{x\}),
\label{timeevol}	
\end{equation}
where $E_{\{q\}}=\sum_{j=1}^N q_j^2/2$. Note that as long as we start with the system ground state $\Psi=\Psi_0$, which possesses a zero total momentum, the only contributions in the expansion in Eq.~(\ref{timeevol}) are related to eigenstates $\widetilde{\Psi}_{\{q\}}$ with $P=\sum_{j=1}^N q_j=0$. In addition, we restrict the expansion to the the eigenstates $\widetilde{\Psi}_{\{q\}}$ corresponding to non-negligible amplitudes $\alpha_{\{q\}}$. In Tab.~\ref{tab::overlapy} we present values of leading amplitudes obtained with the help of the exact formula (\ref{eqn::overlap}) for $N=4$.

 The analytical results of the overlaps can be also used as a benchmark for the Monte-Carlo integration employed by us to calculate $\rho_{\rm rel}(x;t), \, G^{(2)}(x-y;t)$ and $P(x,y;t)$. Within the Monte Carlo approach the overlaps can be calculated by means of the following summation.
\begin{equation}
\big<\widetilde{\Psi}_{\{q\}}\big|\Psi_0 \big> \approx \frac{1}{N_\mathcal{U}}\sum_{\{x\} \in \mathcal{U}} \widetilde{\Psi}^*_{\{q\}}(\{x\})\Psi_{0}(\{x\}),
\label{eqn::overlapMC}
\end{equation}
where $\mathcal{U}$ is a collection of $N_\mathcal{U}$ sets of positions $\{x\}$ randomly chosen from the uniform distribution. In our simulations we take $N_\mathcal{U}=10^6$ sets of positions, which allows us to determine the overlaps with accuracy of the order of $10^{-3}$, see Tab.~\ref{tab::overlapy}. The integrals in Eqs.~(\ref{rho1mod}), (\ref{g2}) and (\ref{Pxyt}), can be computed in a similar way as the overlaps in Eq.~(\ref{eqn::overlapMC}). The overall overlap of eigenstates used for obtaining $\rho_{\rm rel}(x;t), \, G^{(2)}(x-y;t)$ and $P(x,y;t)$ equals to $0.9946$ and $0.9940$ for Monte Carlo integration and analytic one, respectively.
\\

\begin{table}[H]
\center
\begin{tabular}{c|r|c|c}

 $\{q\}/(2 \pi)$  & energy & $|\alpha_{\{q\}}|^2$ & $|\alpha_{\{q\}}|^2$ \\ 
 & &  Eq.~(\ref{eqn::overlap}) & Eq.~(\ref{eqn::overlapMC})\\ \hline\hline
-0.003 , -2.52i, 2.52i,  0.003  & -249.7   & 0.458422 & 0.458319    \\ \hline
-3.77i, -1.26i, 1.26i, 3.77i   & -624.1  & 0.227098 & 0.228112    \\ \hline
-0.37-1.27i, -0.37+1.27i,  & -115.5   & 0.068781 & 0.068779    \\
 0.37-1.27i,  0.37+1.27i & & &\\ \hline
-0.43-2.52i, -0.43+2.52i,  & -205.4   & 0.053255 & 0.053224    \\ 
-0.43, 1.30 & & &\\ \hline
-1.30 ,  0.43,  & -205.4  & 0.053255 & 0.053224    \\ 
0.43-2.52i, 0.43+2.52i & & & \\ \hline
1.02, -1.26i, 1.26i, -1.02    & -22.5    & 0.050643 & 0.050505    \\ \hline
-1.03+1.26i, -1.03-1.26i,  & -40.6    & 0.024169 & 0.024250   \\ 
1.03+1.26i, 1.03-1.26i & & & \\ \hline
-2.37, -1.26i, 1.26i, 2.37   & 158.7    & 0.011749 & 0.011761    \\ \hline
-2.48, 0.83+2.52i,  & -88.1   & 0.006947 & 0.006761  \\ 
0.83, 0.83-2.52i & & &  \\ \hline
-0.83+2.52i, -0.83, & -88.1   & 0.006947 & 0.006761  \\ 
-0.83-2.52i, 2.48 & & &  \\ 
\end{tabular}
\caption{ Quasimomenta $\{q\}$ divided by $2 \pi$, energies and probabilities $|\alpha_{\{q\}}|^2=\big|\big<\widetilde{\Psi}_{\{q\}}\big|\Psi_0 \big>\big|^2$ corresponding to relevant eigenstates $\widetilde{\Psi}_{\{q\}}$ of the 4-particle system after the fourfold quench of the interaction strength from $g_0=-3.95$ to $\widetilde{g}_0=4g_0$. The quasimomenta $\{q\}$ which determine the eigenstates $\widetilde{\Psi}_{\{q\}}$ were obtained as described in Appendix~\ref{AA}, starting with different sets of integers $\{d_1,\ldots,d_4\}$. The analytical results for overlaps $\alpha_{\{q\}}$, Eq.~(\ref{eqn::overlap}), are compared with the results obtained by employing the Monte Carlo integration, Eq.~(\ref{eqn::overlapMC}). The values $\{q\}/(2\pi)$ are rounded to $10^{-2}$ for real and imaginary part.}
\label{tab::overlapy}
\end{table}

\bibliography{ref_breather} 


\end{document}